\documentstyle[psfig,11pt,aas2pp4,amssym]{article}

\begin{document}

\def\refindent{\par\raggedright\noindent\parskip=2pt\hangindent=3pc
  \hangafter=1}

\def\refp#1#2#3#4{{\refindent{#1.}{ {#2}, }{{\bf#3}}{, #4.}\par}}

\def\refb#1#2#3{{\refindent{#1}{ {\it#2}, }{#3.}\par}}

\def\refx#1{{\refindent{#1}\par}}

\def\mpc{\rm{\ Mpc}}
\def\kpc{\rm{\ kpc}}
\def\kms{\rm{\ kms}^{-1}}
\def\msun{\rm{\ M}_\odot}
\def\ie{{\it i.e.}}
\def\eg{{\it e.g.}}
\def\etal{{\it et al.}}

\font\tfont=cmbx10                         
\font\ttfont=cmbx7                         
\font\tttfont=cmbx5                        
\newfam\vecfam
\def\vecfont{\fam\vecfam\tfont}
\textfont\vecfam=\tfont \scriptfont\vecfam=\ttfont
\scriptscriptfont\vecfam=\tttfont
\def\b#1{{\vecfont #1}}

\title{Time stepping N-body simulations} 

\author{Thomas Quinn,
Neal Katz\altaffilmark{1},
Joachim Stadel, and  George Lake
}
\affil{Astronomy, Box 351580, University of Washington, Seattle, WA  98195}

\altaffiltext{1}{Current address:
Department of Physics and Astronomy
517 Lederle Graduate Research Tower
University of Massachusetts
Amherst, MA  01003-4525}


\begin{abstract}

  Leapfrog integration has been the method of choice in N-body
  simulations owing to its low computational cost for a symplectic
  integrator with second order accuracy.  We introduce a new leapfrog
  integrator that allows for variable timesteps for {\em each}
  particle in large N-body simulations.  Tests with single particles
  in fixed potentials show that it behaves as a symplectic integrator.
  We then examine the results of both standard leapfrog and our
  temporally adaptive leapfrog on full N-body integrations of clusters
  and large scale structure establishing accuracy criteria for both
  methods.  The adaptive method shows significant speed-ups over
  single step integrations---but the integrator no longer appears to
  be symplectic or, in the case of large scale structure simulations,
  accurate.  This loss of accuracy appears to be caused by the way
  that the timestep is chosen, not by the integrator itself.  We
  present a related integration technique that does
  retain sufficient accuracy.  Although it is not symplectic, it is apparently
  better than previous implementations and is our current integrator
  of choice for large astrophysical simulations.  We also note that
  the standard leapfrog difference equations used in cosmological
  N-body integrations in comoving coordinates are not symplectic. We derive an
  implementation of leapfrog that is in comoving canonical coordinates
  to correct for this deficiency.

\end{abstract}


\section{Introduction}

Over the past decade, spatially adaptive methods have been developed
to calculate gravitational forces in N-body simulations.  These
include tree codes (Appel, 1985; Barnes and Hut, 1986), fast multipole
methods (Greengard, 1987), and adaptive grids  (Couchman, 1991). 
As the number of 
particles in an N-body simulation
grows, so do the density contrasts.
Hierarchical methods can follow
extremely large dynamic ranges in densities at modest additional
cost per force evaluation.  However, large
ranges in densities also imply a large range in 
time scales ($\propto 1/\sqrt{\rm density}$).
If we take the final
state of a simulation and weight the computational work done on particles
not uniformly but
inversely with their natural timesteps, we find a potential gain of $\sim 50$.  
Temporal adaptivity is one of the last algorithmic areas where we can
target an order of magnitude improvement.
Hence, we seek a hierarchical integrator, \ie, 
a method such that particles are on adjustable individual
timesteps.

The most commonly used time integration scheme for N-body simulations
is the leapfrog method.  Leapfrog has several advantages over other
methods.  1) For second order accuracy only one force evaluation and
one copy of the physical state of the system is
required.  This is particularly beneficial for N-body simulations where
the cost of a force evaluation is very expensive.  2)  The force
field in an N-body simulation is not very smooth, so higher order does
not necessarily mean higher accuracy.
3)  It is a symplectic integrator, \ie, it
preserves properties specific to Hamiltonian systems.  See Channell
and Scovel (1990) for a review of symplectic integrators.
Gravitational N-body systems are Hamiltonian, and therefore they
should benefit from the use of an integrator that conserves phase
space volume and has no spurious dissipation.  This could be
especially important in self-gravitating systems where a dissipation
time scale that is linked to the dynamical time can lead to a runaway
to spuriously large densities.
{\em Leapfrog is a second order symplectic integrator requiring only one costly
force evaluation per timestep and only one copy of the physical state of the 
system.}
These properties are so desirable that we concentrate 
on making an adaptive leapfrog.

Hierarchical leapfrogs have been used
before (Porter 1985; Ewell 1988; Hernquist and Katz 1989),
but they were not 
symplectic (\S 4).   There exist symplectic 
integrators with individual but {\em fixed} timesteps for either particles or modes
(Saha \& Tremaine 1994; Skeel \& Biesiadecki 1994;
Lee, Duncan \& Levison 1997).
Hut, Makino \&
McMillan (1995) proposed an iterative scheme for choosing 
timesteps for a single particle in a rapidly varying potential, but 
each iteration involves a force evaluation that is
prohibitively expensive for large N simulations (\S 2).

Section 2 describes the theory behind a hierarchical 
integrator.
Section 3 describes tests of this integrator on a single particle
in a potential, and Section 4 will present tests of the integrator on
full N-body systems.  We discuss the implications of these
results in Section 5.

\section{Symplectic Integrators}

A symplectic integrator is an exact solution to
a discrete Hamiltonian
system that is close to the continuum 
Hamiltonian of interest.  Therefore, it
preserves all the Poincar\'e invariants and  places
stringent conditions on the global geometry of the dynamics.  An
obvious example is total energy conservation in a system with a
time-independent Hamiltonian or the conservation
of angular momentum in axisymmetric systems
(Zhang \& Skeel 1995).  A symplectic integrator will exactly conserve
the energy in the discrete Hamiltonian that is
an approximation to the true energy of the system.  This approximate
energy oscillates about the true energy without any 
numerical dissipation.  

\begin{figure}[tb]
\psfig{figure=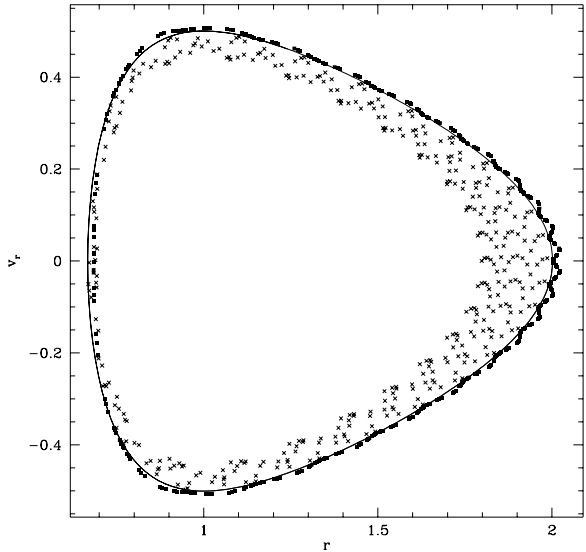,width=\hsize}
\caption{A comparison of symplectic and non-symplectic integrators is
  made.  The squares are a second order leapfrog integrator; the
  crosses are a 4th order Runge-Kutta integrator with the same
  timestep, and the solid line is the exact solution.}
\label{phase}
\end{figure}

The difference between the discrete and continuum Hamiltonians
can be viewed as a small 
perturbation given by the truncation error of the
integrator.  The error is a Hamiltonian!
If the error Hamiltonian is a 
sufficiently small perturbation, then the KAM theorem (Arnold
1978) guarantees that the invariant curves destroyed are a
set of finite measure.  In other words, almost all orbits that are
stable in the real system will continue to be stable in the numerical
system.  An illustration of these advantages is shown in Figure
\ref{phase}.  Here the radial velocity, $v_r$, is plotted against the radius,
$r$, for an ellipticity, $e = 0.5$
Kepler orbit using a leapfrog integrator and using a fourth order Runge-Kutta
integrator.  In each integration,
approximately 24 steps were taken per orbit, and the integrations ran
for 16 orbits.  Note how the leapfrog integrator oscillates about the
true solution but always remains on a one dimensional surface.
This indicates that it
is indeed conserving an energy-like quantity, \ie\ having the orbit
constrained to a one dimensional surface shows the existence of an isolating
integral of motion.  On the other hand, the
Runge-Kutta orbit slowly becomes more circular.  The poor performance
of the Runge-Kutta integrator is remarkable given that it is a fourth
order integrator and uses four times as many force evaluations as
the leapfrog integrator.  Also note the large wiggles in the leapfrog
integration at apoapse.  These are indicative of the proximity of
resonant islands that would lead to a instability for larger timesteps.

The leapfrog integrator can be written as
$$
\begin{array}{ll}
\b r_{n+1/2} &= \b r_n + {\textstyle {1\over 2}} \tau \b v_n, \\
\b v_{n+1} &= \b v_n + \tau \b a(\b r_{n+1/2}), \\
\b r_{n+1} &= \b r_{n+1/2} + {\textstyle {1\over 2}} \tau \b v_{n+1},
\end{array}
$$
where $\b r$ is the position vector of a particle, $\b v$ is the
velocity, $\b a$ is the acceleration, and $\tau$ is the timestep.
When several of these steps are put together, the two position updates
can be combined to a single update $\b r_{n-1/2}$ to $\b r_{n+1/2}$,
and the resulting alternation between updating $\b r$ and $\b v$
gives leapfrog its name.  The symplectic nature of leapfrog can be
seen by noting that the position update is equivalent to evolving the
system exactly under the Hamiltonian $H_D = {1\over 2} v^2$, and the
velocity update is equivalent to evolving the system under the
Hamiltonian $H_K = V(\b r)$, where $V(\b r)$ is the potential
generating the accelerations.  We
will call the operator that evolves the system under $H_D$ the
``drift'' operator, $D$, and the operator that evolves the system under
$H_K$ the ``kick'' operator, $K$.\footnote{
If one expresses Hamilton's equations as $\dot\b z = \{ \b z, H \}$,
where $\b z$ are the phase space coordinates and $\{\ ,\ \}$ are
Poisson brackets, then formally $D(\tau)\b z = \exp(\tau \{\b z,
H_D\})$, and $K(\tau)\b z = \exp(\tau \{\b z, H_K \})$.
}
It can be shown that (see Saha and
Tremaine, 1992 for details) the combination of operators
$D(\tau/2)K(\tau)D(\tau/2)$ evolves the system under a Hamiltonian
$$
H_N = H_D + H_K + H_{err} = {\textstyle 1\over 2} \b v^2 + V(\b r) +
H_{err},
$$
where $H_{err}$ is of order $\tau^2$.  The existence of this surrogate
Hamiltonian ensures that the leapfrog is symplectic and second order.
Since the Hamiltonian is symmetric with respect to $H_D$ and $H_K$,
the combination of operators $K(\tau/2)D(\tau)K(\tau/2)$ is also a
symplectic second order integrator.  Explicitly, this is
$$
\begin{array}{ll}
\b v_{n+1/2} &= \b v_n + {\textstyle {1\over 2}} \tau \b a(\b r_n), \\
\b r_{n+1} &= \b r_n + \tau \b v_{n+1/2}, \\
\b v_{n+1} &= \b v_{n+1/2} + {\textstyle {1\over 2}} \tau \b a(\b r_{n+1}).
\end{array}
$$
Higher order symplectic integrators can be constructed from
combinations of leapfrog steps (Yoshida 1990).
Each $N$th order leapfrog integrator requires $N-1$ force evaluations
and only one copy of the physical state of the system.

Unfortunately, constructing a variable step-size method by choosing a
new timestep after each leapfrog step gives very disappointing results
(Calvo and Sanz-Serna, 1993).  Some simple schemes can be shown to
have no rigorous stability criterion for any step size
(Skeel 1993). 
Several explanations have been given for
this behavior.  The simplest is to note that a variable step integrator
is evolving a dynamic system with state variables ($\b r$, $\b v$, $\tau$).
The projection onto the phase space coordinates ($\b r$, $\b v$) cannot
be described by a Hamiltonian.  
Skeel and Gear (1992) consider
a one step symplectic operator of the form
$$
(\b r_{n+1}, \b v_{n+1}) = F(\b r_n, \b v_n, \tau(\b r_n, \b v_n)),
$$
and show that if $F$ is symplectic for a constant step size, it will
not in general be symplectic for variable step size.  Another way to
see this problem is to notice that the time reversibility has been broken:
if we step forward in time and then step backward, we do not end up at
the same point because of the change in timestep.  Since symplectic
implies time reversible, such an operator is not symplectic.
Integrators that are time reversible are referred to as
reflexive by Kahan (1993;
see also Kahan and Li 1997)
who argues that reflexivity is the key to the
robust properties of ``updating formulae" rather than 
symplecticity.

A strategy to make 
leapfrog reflexive are clear from a cursory look at its 
evolution operators.   The operation
$D(\tau/2)K(\tau)D(\tau/2)$ is reversible if we use an 
``select'' operator to choose the timestep, $S$, such
that the time
reversibility is retained.  Hut, Makino, and
McMillan (1995) achieve this by 
using an implicit definition of the timestep:
$$
\tau = {\textstyle 1 \over 2} [ \tau(\b r_n, \b v_n) + \tau(\b
r_{n+1}, \b v_{n+1})].
$$
So, the beginning and end of the step are required to agree on the timestep.
One can solve for $\tau$ iteratively, and very good results are
obtained even with only one or two iterations.  (This result
is also clear from Kahan 1993).
This iterative scheme poses several problems when applied to a large
N-body simulation.  
It requires backing up timesteps,
throwing away  expensive force calculations,  using
auxiliary storage and must specify a method
for synchronizing the particles for mutual force evaluations.  

However, we can also see an alternative means to restore reflexivity.
Let us choose a select operator $S$ that commutes with
$K$, so that $DSKD$ is equivalent to $DKSD$.  Since $K$ only changes
the velocities and not the positions, an $S$ operator that depends
entirely on positions satisfies the commutation requirement.  As we
shall show below, this is not strictly time reversible.  However, since the
operators read the same forward and backward, we will refer to $DSKD$ as a
``palindromic'' integrator.  Synchronization can be maintained by only
choosing timesteps that are a power-of-two subdivision of the largest
timestep, $\tau_s$.  That is,
$$
\tau_i = {\tau_s \over 2^{n_i}},
$$
where $\tau_i$ is the timestep of a given particle, and $n_i$ is an integer
(See Hernquist and Katz, 1989).  Combining this
synchronization procedure with the $DSKD$ evolution operation we have
the following (recursive) algorithm for a timestep. 1) Drift the
particles forward $\tau_i/2$.  2) Apply the select operator.  If it
accepts this step, then finish the step with $K(\tau_i)$ and
$D(\tau_i/2)$.  Otherwise, drift the particles back $\tau_i/2$ and
take two timesteps using $\tau_i/2$.  The algorithm can  be
expressed recursively in the following pseudo-code:
\vbox{
{\obeylines
\tt
Timestep($\tau$)
$\{$
\quad Drift($\tau$);
\quad Select($\tau$);
\quad {\bf if} $\tau$ = $\tau_{min}$
\quad $\{$
\qquad Kick($\tau$);
\qquad Drift($\tau$);
\quad $\}$
\quad {\bf else}
\quad $\{$
\qquad Drift(-$\tau$);
\qquad Timestep($\tau$/2);
\qquad Kick($\tau$);
\qquad Timestep($\tau$/2);
\quad $\}$
$\}$
}
}

Here, {\tt Drift($\tau$)} applies $D(\tau)$ to all particles,  {\tt
Kick($\tau$)} applies $K(\tau)$ to particles with timestep $\tau$ and
{\tt Select()} decides to which timestep the particles belong, with
the side effect of finding the minimum timestep, $\tau_{min}$.  We
will refer to the traditional method of choosing a timestep at the
beginning of the integration step as $SDKD$.

\begin{figure}[tb]
\psfig{figure=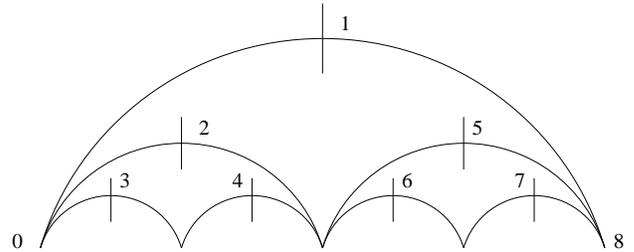,width=\hsize}
\caption{Diagram of particles on three
separate timesteps, where arcs represent ``drifts'', and vertical
lines represent ``kicks''.  For $DSKD$, the order of flow throw the
diagram is 0, 1, 2, 3, 2, 4, 1, 5, 6, 5, 7, 8.}
\label{umbrella}
\end{figure}

As an example of how this procedure works, consider
Figure \ref{umbrella}.  In this diagram we consider the case where there
are particles on three separate timesteps.  The arcs represent
drifting the particles, and the vertical slashes represent
either an select or kick of particles at that timestep.  Starting at
point 0, all particles are drifted to point 1 and evaluated for
whether they are on the largest timestep.  Then all particles are
drifted to point 2 and those particles not on the largest timestep are
evaluated for whether they are on the middle timestep.  Then all
particles are drifted to point 3 where those particles on the smallest
timestep are kicked.  All particles are then drifted to point 2 where
particles on the middle timestep are kicked, then to point 4 where
particles on the smallest timestep are kicked, then to point 1 where
particles on the largest timestep are kicked.   Then all particles are
drifted to point 5 where again all particles not on the largest
timestep are evaluated for whether they are on the middle timestep.
Then the appropriate particles are kicked at points 6, 5, and 7.
Finally all particles are drifted to point 8 and a single large
timestep is complete.

\begin{figure}[tb]
\psfig{figure=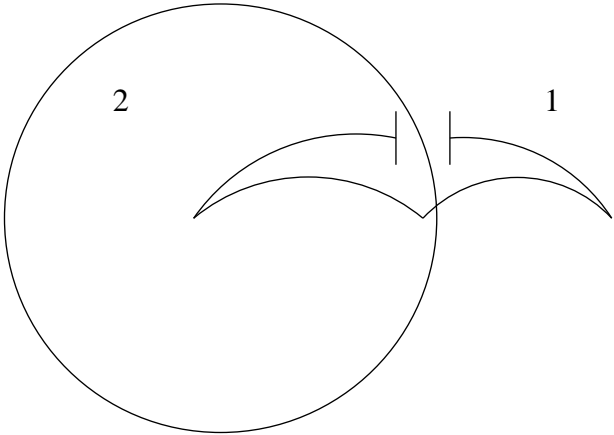,width=\hsize}
\caption{An example of a non-time-symmetric scenario is shown.  If the
step starts in region 2, two steps are taken.  If the step starts in
region 1, only one step is taken.}
\label{Oops}
\end{figure}

There are several things to note about the above algorithm.  One is
that it
is not unique.  It uses a ``top down'' approach by trying the
largest timestep and reducing it until it is deemed appropriate.  A
``bottom up'' scheme can be implemented where a smallest timestep is
tried first.  Secondly, this algorithm is not exactly time reversible.
Scenarios can easily be constructed where a forward step followed by a
backward step will not come back to the initial conditions.  An
example is shown in Figure \ref{Oops}.  The circle delineates the
boundary between region 1, where one timestep is needed, and region 2,
where a timestep of 1/2 the timestep in region 1 is needed.  If we
start a particle in region 2, it is drifted forward, found to still be
in region 2, so two timesteps are taken.  If we start the particle in
region 1 and go backward, it is possible that the drift leaves the
particle still in region 1, so only one timestep is taken going
backward, and we do not arrive at the point from which we started. Finally,
since several selects may be done for every kick, the efficiency of
this algorithm depends upon {\tt Select} being able
to quickly decide an appropriate timestep given the particle
positions.  A suitable criterion that depends only on the positions is
one based upon the local dynamical time $t_d = 1/\sqrt{G\rho}$.  With
this criterion, {\tt Select} will pick the largest timestep $\tau$
such that
\begin{equation}
\label{dencrit}
\tau < {\eta \over \sqrt{G\rho}}, 
\end{equation}
where $\eta$ is a constant to be determined based on stability and
accuracy requirements.  With this criterion, region 2 in Figure \ref{Oops} is
a high density region while region 1 has low density.  Similar
problems to this loss of reflexivity
can occur in the iterative schemes (Hut, Makino, \& McMillan 1995; Kahan 1993)
if the starting guess of an iteration
is preconditioned by the history of the particle.  That is, the loss of
reflexivity depends as much on the choice of a ``top down" approach
as on the {\tt Select} operator.


\section{Single Particle Tests}


Our first test is the motion of a
test particle in the one dimensional effective potential of the Kepler
problem.  If the integrator performs well, these
tests will  guide our choice of $\eta$ for later
applications.  The timestep criterion in equation \ref{dencrit} is
not straight-forward to implement in this case, since the
density is zero anywhere outside the central source.  
Therefore, in this test we
determine the timestep by the enclosed density, $\rho_e = 3M/4\pi r^3$
where $M$ is the mass of the central object.

\begin{figure}[tb]
\psfig{figure=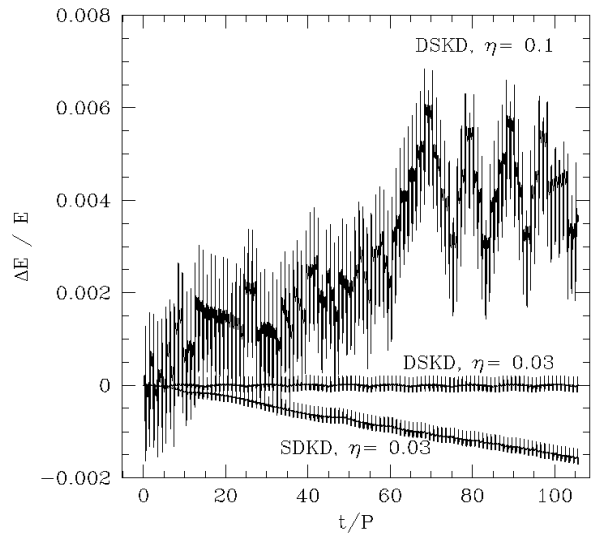,width=\hsize}
\caption{Relative change in total energy is
plotted as a function of time in units of the orbital period for
several integrators.  Curves are plotted for $DSKD$ with $\eta = 0.1$
and $\eta = 0.03$, and $SDKD$ with $\eta = 0.03$.}
\label{keplere}
\end{figure}

Figure \ref{keplere} shows the relative change in total energy as
function of time for several
integrators.   The orbit has eccentricity $e = 0.5$, and is evolved
for about 100 periods.  Results for two values of $\eta$ for the $DSKD$
method are plotted as well as a result for $SDKD$.  A fixed step
integration was also run, and its energy nearly coincides with the
$\eta = .03$ $DSKD$ method.  From the energy plot,
it appears that the $DSKD$ method with $\eta =
0.03$ is indeed symplectic.  In contrast, the $SDKD$ method with the
same timestep criterion experiences a secular drift in the energy.
The savings in force evaluations is also significant.  A fixed
timestep integration with similar energy errors required 50000 force
evaluations to accomplish 100 orbits
while the adaptive timestep integration required less than 16000,
a savings of over a factor of 3 for this moderate
eccentricity orbit.  Also, it can be seen from the energy plot that
$\eta \approx 0.03$ is needed for the adaptive algorithm to be
stable.  This is because a stable orbit, having one integral of motion
for each degree of freedom, should be quasiperiodic, which the $\eta =
0.03$ line appears to be.\footnote{Remember that if there is one
integral for each degree of freedom, the Hamiltonian can be expressed
as $H = H(\b J)$, and any function of the phase space coordinates,
$f(\b J, \b \theta)$ can be expressed as $f(t) = f(\b J, \b \omega t)$
where $\b J$ and $\b \omega = \partial H/\partial \b J$ are constants.}
For $\eta$ larger than this, the method does not appear to be
symplectic, but the changes in energy appear to be due to instability,
(\ie \ integrals being destroyed)
rather than a secular buildup of truncation error.


\begin{figure}[tb]
\psfig{figure=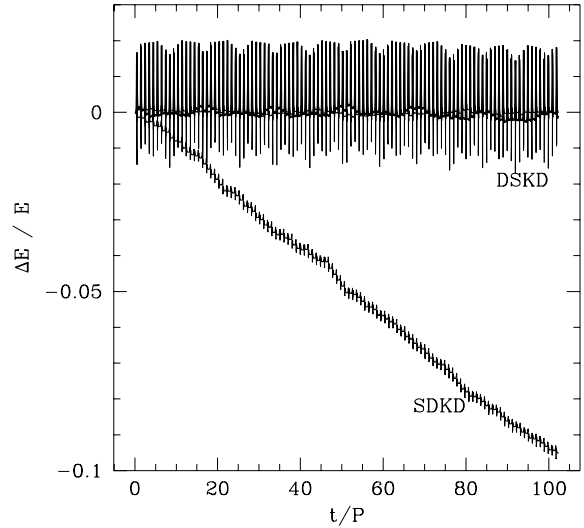,width=\hsize}
\caption{  Relative change in total energy is
plotted as a function of time in units of the orbital period for
several integrators in an isothermal potential.  Curves are plotted
for $DSKD$ with $\eta = 0.1$ and $\eta = 0.03$, $SDKD$ with $\eta = 0.03$.}
\label{isoe}
\end{figure}

A more realistic test of the integrator is the motion of a single
particle in the potential of an isothermal sphere.  As well as having
a density defined everywhere, allowing the use of the local density in the
timestep criterion, the potential is similar to that of 
globular clusters, galaxies and clusters of galaxies
in our large N-body simulations.
Figure \ref{isoe} shows the energy conservation for the $SDKD$ and $DSKD$
methods for a particle in the potential of a singular isothermal
sphere.  The ratio of apocenter to pericenter is 3.2:1.  The density
used for choosing the timestep is the local density of the isothermal
sphere.  Again the $SDKD$ shows a secular drift in the energy, while the
$DSKD$ has the characteristics of a symplectic method.  Also note that in
the case of an isothermal sphere, the method is stable for $\eta =
0.1$.

\section{N-Body Tests}

For our purposes, the real proof of a given integration method is how well it performs in an
actual N-body code.  In this section we test the above methods in two
complementary systems.   The first is a King model---a model which, in
the absence of collisional relaxation, should remain static.  The
second is a cosmological simulation of a universe dominated by cold
dark matter (CDM).  This case is very dynamic, with structure evolving
on all scales.   All the simulations below were performed with {\em
PKDGRAV}  (Dikaiakos \& Stadel, 1996; Stadel \& Quinn, in
preparation), a parallel code that
uses a binary tree with a Barnes and Hut (1986) style opening
criterion to make the gravity calculation order $N\log(N)$.  To
mitigate the effect of force errors, we use an opening
criterion $\theta=0.55$, and accelerations from cells are expanded
to hexadecapole order in all the following simulations.


Before we test the multistep model in a full N-body simulation, let us
explore the more basic question of what single fixed
timestep is appropriate for an astronomically interesting
N-body model.
For
our tests we will use a $W=9$ King model realized with 100,000
particles evolved for 100 central
dynamical times.  This is representative of a galaxy halo over its
lifetime in a typical cosmological N-body simulation.

\begin{figure}[tb]
\psfig{figure=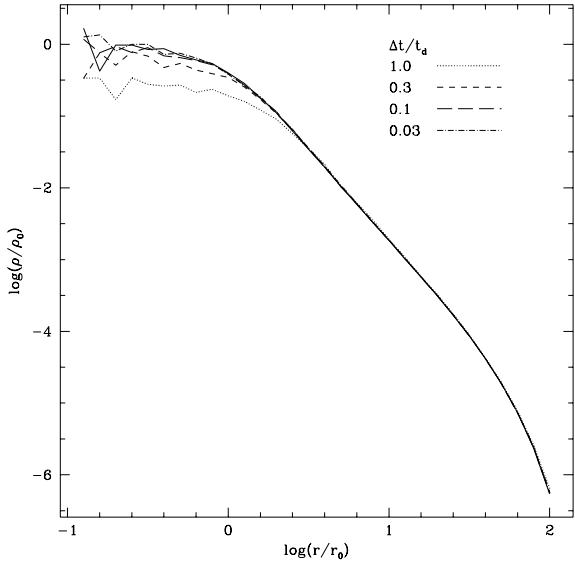,width=\hsize}
\caption{The spherically
averaged density in
units of the central density is plotted against radius in units of the
core radius for the end
state of several simulations.  The solid line is the density profile
of the initial King model.  All integrations were done with a fixed
step leapfrog integrator with the timestep in units of the central
dynamical time given in the legend.
\label{leapfden}
}
\end{figure}

\begin{figure}[tb]
\psfig{figure=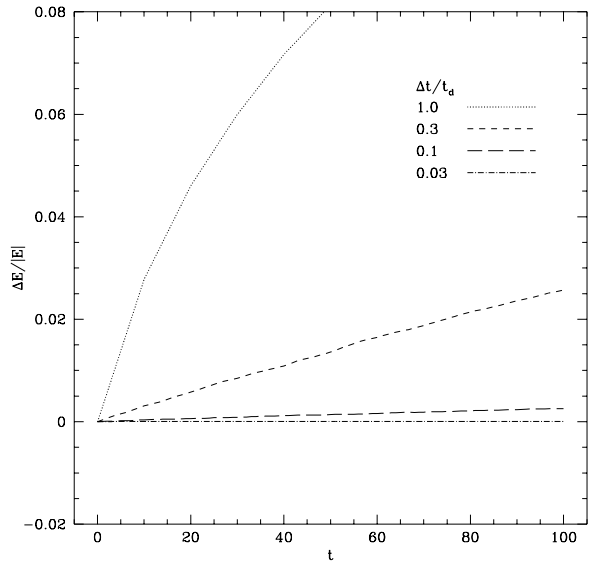,width=\hsize}
\caption{The relative change in the total
energy as a function of
time for the same simulations as in Figure \protect\ref{leapfden}.
The time is in units of the central dynamical time.
}
\label{leapfeng}
\end{figure}

Energy conservation is a typical measure of the quality of a
simulation, but more appropriate measures should involve the
convergence of scientifically interesting quantities.  For the King
model tests we will use the radial density profile which should remain
constant in a collisionless simulation.  In Figure \ref{leapfden} we
compare the density profile of the initial King model with that at the
end of a simulation with various timestep sizes evolved over
100 central crossing times.
All models were integrated with fixed step leapfrog using
timesteps of either 1.0, 0.3, 0.1, or 0.03 times the central dynamical
time.  From the figure, we see that $\tau < 0.1 t_d$ is needed to
maintain the central density of the King model.
Figure \ref{leapfeng} shows the evolution of the total energy for these
same simulations.  Note that an energy conservation of better than 3\% is
needed to preserve the density profile of this King model.

In order to use a density criterion to determine timesteps in a
general N-body simulation, it is necessary to have a method for
calculating the local density for every particle in an arbitrary
distribution.  We use a Smooth Particle
Hydrodynamics estimate of the density by smoothing over a fixed
number of nearest 
neighbors with a cubic spline kernel.  The smoothing
length, $h$, is adaptive and is set so that there are exactly 64 particles
within $2h$.  The kernel is made symmetric using the
``gather-scatter'' algorithm as described in Hernquist and Katz
(1989).  

\begin{figure}[tb]
\psfig{figure=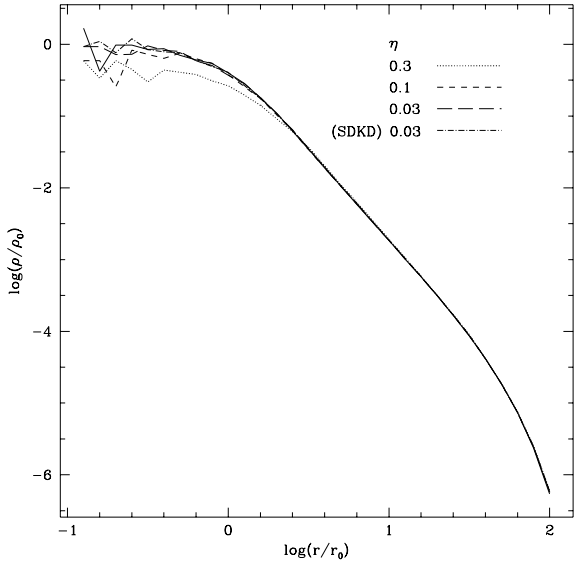,width=\hsize}
\caption{The spherically averaged density is
plotted against radius as in Figure \protect\ref{leapfden}.  The
integrations were
done using a multistep leapfrog with a density timestep
criterion.  All integrations except the one indicated used a $DSKD$
style timestepping.
\label{dakdden}
}
\end{figure}

\begin{figure}[tb]
\psfig{figure=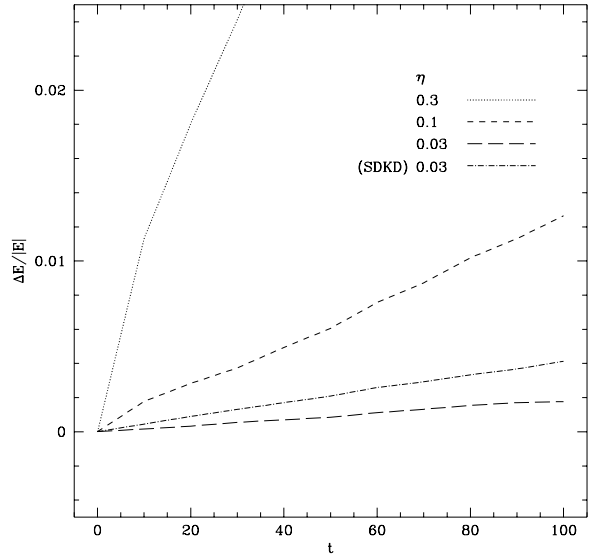,width=\hsize}
\caption{The relative change in the total
energy as a function of
time for the same simulations as in Figure \protect\ref{dakdden}.  The
time is in units of the central dynamical time.
\label{dakdeng}
}
\end{figure}

We show the radial density profiles for adaptive timestep integrations
in Figure \ref{dakdden}.  From the figure, we see that, as in the single
particle case, $\eta \leq 0.03$ is needed to integrate this model with
reasonable accuracy.  The $DSKD$ integration with $\eta =
0.03$ needed a factor of 6 less force evaluations than the fixed step
integration with $\tau/t_d = 0.03$, or a factor of 2 less than the fixed
step, $\tau/t_d = 0.1$ integration.  As shown in Figure \ref{dakdeng}, the
energy conservation of the adaptive scheme is comparable to the fixed
step integrator.  Also shown in the figures are results for the
$SDKD$ integrator.  It appears that at the level of accuracy
needed to integrate this King model for 100 dynamical times, the
secular drifts introduced by the $SDKD$ integrator do not
significantly affect the integration.


%


\def\rprime{{{\b r}^\prime}}
\def\vprime{{{\b v}^\prime}}
\def\pprime{{{\b p}^\prime}}

Before testing timestepping in a cosmological simulation we first note
that the standard leapfrog difference equations used in cosmological
simulations with comoving coordinates are not done in canonical
coordinates and are apparently not symplectic.
However, as shown in the appendix, a symplectic integrator can easily be
derived, for which the ``drift'' and ``kick'' operators are
$$
\begin{array}{l}
D(\tau) \equiv \rprime_{t + \tau} = \rprime_{t} + \pprime \displaystyle
\int_t^{t + \tau} {dt \over a^2} \\
K(\tau) \equiv \pprime_{t + \tau} = \pprime_t - \nabla^\prime
\phi^\prime \displaystyle \int_t^{t+\tau} {dt \over a},
\end{array}
$$
where the variables are as defined in the appendix.
We use this integrator in the tests that follow.
This integrator has another advantage over the standard leapfrog
difference equations in that it can be used to implement both the
$DKD$ and $KDK$ form of leapfrog.  The standard equations can only be
derived for $DKD$.  The $KDK$ form of leapfrog can have advantages
over $DKD$ (see the discussion of costs in \S 5).  

Energy conservation may be a an even poorer criterion for
determining the accuracy of a cosmological integration.  
This owes to the possibility of making small errors in calculating the
energy of the uniform background that 
can dominate the total error budget and
mask any problems in the structure of interest.
Nonetheless, it pays to examine the evolution of the energy.  It was
transients in the energy at the beginning of the cosmological simulations
that made us suspect that the integration was not in canonical coordinates,
leading to the results in Appendix A.   
But, these large transients in energy had little effect on the
three dimensional structure, underscoring our point that it may be
a poor diagnostic.
Small changes in the way that two simulations 
handle global energy may be an interesting clue to other aspects of the
integrator.

Since there is
no analytic model for the full three dimensional formation of non-linear
cosmological structure, we have to use another simulation to determine
the ground truth.
Therefore,  we test the
goodness of a given timestepping algorithm  by examining its
convergence to a
simulation with a large number of timesteps for all particles.
This convergence will be assessed 
based on properties of the groups that are
formed and the internal structure of the largest group.



The fiducial simulation is of a standard CDM dominated
universe in a periodic box of size $22.2 \mpc$ ($h = 0.5$) on a side
realized on a grid of $32^3$ particles.  This implies a
particle mass of $2.3\times 10^{10} \msun$.  A spline softening is used
with a softening length of $20 \kpc$.  The initial conditions were imposed
on the particles at a redshift of 49 using the Zel'dovich
approximation. The small size of this box
means it does not accurately represent large scale structure, but it
does guarantee significant non-linear evolution and a stringent test
for the time integrator.


\begin{figure}[tb]
\psfig{figure=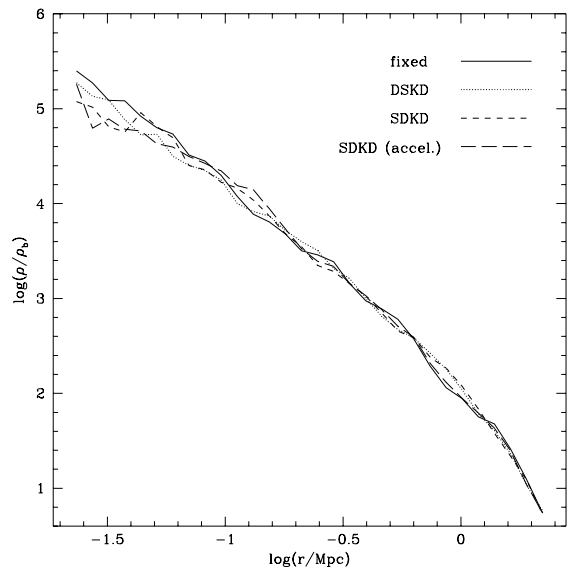,width=\hsize}
\caption{The density averaged in spherical shells is plotted as a
function of radius for the largest halo in the cosmological simulation.
\label{cosmoprof}
}
\end{figure}

The first convergence test is the density profile at the end of the simulation
of the largest
halo as identified using a
friends-of-friends grouping algorithm.
The object in question has a total mass
of $1.23\times 10^{14} \msun$  within the virial radius and a maximum
circular velocity of $639 \kms$.  This test looks at how the
timestep algorithm affects the structure of well resolved objects in
the simulation.  Since structure forms hierarchically, the objects in
the largest clusters will have experienced a complicated history with
several dynamical times in previous generations of structure.  Any
effect that accumulates with dynamical time is most likely to betray
itself in the center of the richest cluster.

Comparisons of the density profile of this object in various
integrations are shown in Figure \ref{cosmoprof}.  The fixed step run
used 4000 steps in a Hubble time.  The $DSKD$ and $SDKD$ runs used $\eta =
0.03$.
As might be expected, the timestep criterion used for the King model
simulation performs well in accurately modeling the density profile of
this cluster-size object.  Looser criteria ($\eta \ge 0.1$)
fail to capture the central density of the object, just as in the
static King model.

\begin{figure}[tb]
\psfig{figure=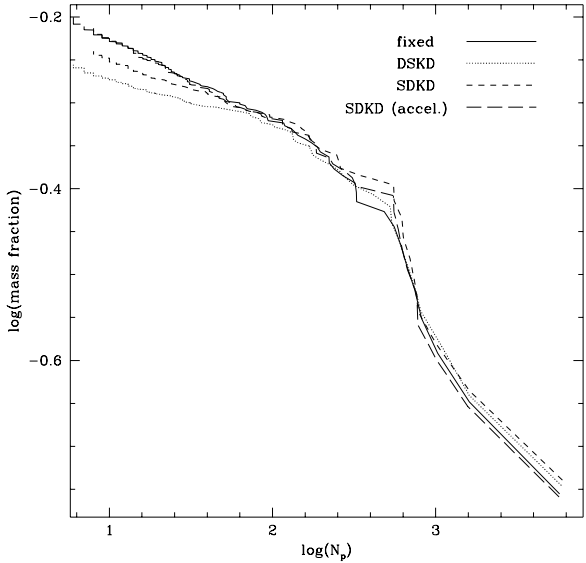,width=\hsize}
\caption{The cumulative mass fraction in groups of a given particle number
or higher is plotted.
\label{cmf}
}
\end{figure}

Our second criterion is the cumulative mass fraction in groups, also
as identified by the friends-of-friends algorithm
with a linking length of 0.26
times the mean interparticle separation.  This linking length
corresponds to an enclosed density equal to the virial density
assuming an isothermal sphere.
Here the
motivation is to see the effect of the timestep algorithm on the
smallest objects we can resolve.  The cumulative mass fraction of
halos is plotted in Figure \ref{cmf}.

For the smallest objects, the density criterion does not do well.
There is a significant decrease in the number of objects formed with
less than one hundred particles in the simulation using the density
timestep criterion.   It may not be surprising that particles residing
in halos with comparable or fewer particles than that used by the smoothing
kernel are unable to accurately estimate their local density.
On the other hand, a simulation using a (non-time
symmetric) multi-stepping algorithm based on the acceleration of the
particles reproduces the numbers of objects seen in the fixed step
simulation all the way down to 8 particles.  (The discrepancy seen at
about 500 particles per group is due to a merger occurring at the end
of the simulation.)

%

\section{Discussion}

First, we examine our results in terms of the number of timesteps
needed to accomplish a given simulation with reasonable accuracy.  
We will determine the number of fixed steps before looking at the
gain from hierarchical timestepping.  For
an equilibrium model where we wish to maintain the overall
density profile, the answer is straightforward: the timestep should be
set to $0.03/\sqrt{G\rho_{max}}$, where $\rho_{max}$ is the maximum
density.  For a cosmological simulation, more assumptions need to be
made.  Consider a simulation in which the largest halo has a circular
velocity, $v_c$, that is roughly constant with radius down to some
core radius, $r_0$.  If we wish to accurately model this halo, then
our timestep criterion must be
$$
\Delta t = {\eta \over \sqrt{G \rho_0}} = {\eta r_0 \over v_c} \sqrt{4
\pi \over 3},
$$
where $\rho_0$ is the (roughly constant) central density, and we have
assumed spherical symmetry.  Now consider a simulation with
a halo with $v_c = 1000 \kms$ where our goal is to  resolve its structure
within 
$10 \kpc$.  In units of the Hubble time, $1/H_0$, we have
$$
H_0 \Delta t \approx 1\times 10^{-3} \eta \left({r_0 \over
10\kpc}\right) \left({v_c \over 1000\kms}\right)^{-1}.
$$
For $\eta = 0.03$, this gives 33,000 steps per Hubble time.  Previous
estimates of the timestep conclude that $6000 (10\kpc/\epsilon)$ timesteps
are needed per Hubble time, where $\epsilon$ is the gravitational
softening length (Lake \etal\ 1995).  This is consistent with our
estimate if we assume that the gravitational softening is 1/5 the core
radius we wish to resolve.

To evaluate the performance of a multistep integrator, we 
must determine the ``fixed
costs'' required for any level of the timestep hierarchy as well
as the costs in determining the timestep level assigned to
each particle.  The total cost of force evaluations is then added
and amortized over the longest timestep.
For example, a tree code requires that 
the tree is built for all the particles
regardless of the number of particles requiring  force evaluations.  
Indeed, for a standard particle mesh code, the forces at every
grid point are either calculated or not.  The only part of the
force evaluation that is not part of the fixed cost would be the
trivial interpolation of the force onto particles.  So, essentially
all the costs are fixed and no gain is possible.

%
%
%

First, we consider the costs in the $SDKD$ scheme with our tree code,
{\em PKDGRAV}, and assume that the cost of actually moving the particles and
evaluating the timestep is negligible.  A single base step with $r$
different timesteps separated by a factor of two, will have a cost of
$$
(2^r - 1)C_t + C_f\sum_{i=1}^r 2^{i-1} N_i,
$$
where $C_t$ is the cost to build the tree, $C_f$ is the cost of
calculating the force on a single particle,  and $N_i$ is the number
of particles on a timestep $i$ of size $2^{1-i}$ times the base step.
For a fixed timestep simulation with all particles at the smallest
timestep, the cost would be
$$
2^{r-1} (C_t + N C_f).
$$
Comparing these two costs, it is easy to see that if $C_t$ dominates
the computation, then in the limit of large $r$, the $SDKD$ scheme would
a factor of two {\em more} expensive than the single step
calculation.  On the other hand, if $C_t$ is negligible, then the
maximum speedup would be a factor of $2^{r-1}$ if almost all the
particles were on the largest timestep.  For a given ratio, $f$, between
$C_t$ and the cost of evaluating all the forces, $N C_f$, there is a
maximum speedup of $(f + 1)/2f$ for large $r$, and almost all of the
particles on the largest timestep.  Note that for an $SKDK$ scheme this
maximum speedup would be a factor of two larger since gravity
evaluations for larger timesteps are synchronized with those of smaller
timesteps, and the same tree can be used for both.  For {\em PKDGRAV} on the
King model with 100,000 particles, the ratio $f$ is about $0.024$, giving
a maximum speedup of about 22.

For the symplectic scheme with a density criterion, the cost of
evaluating the timestep is
non-negligible, and we must account for it.  It also has a fixed cost
part for building the tree, $C_{st}$, and a per particle cost,
$C_{sf}$.  In terms of these costs the total cost of a base step is now
$$
(2^r - 1)(C_t + C_{st}) + C_f\sum_{i=1}^r 2^{i-1} N_i
+ C_{sf}\sum_{i=1}^r (2^i - 1) N_i.
$$
If the tree build costs and per particle costs for selecting the timestep
and evaluating the forces
are similar, then the symplectic scheme would be a factor
of two more expensive than the non-symplectic scheme.   For {\em PKDGRAV}
running on a 100,000 particle King model, the per particle cost of the
density criterion is a factor of 19 smaller than the per particle cost
of a force evaluation, and the tree build for a density tree is 50\%
faster than for a gravity tree.  The maximum speedup assuming an
optimum particle distribution is then about 13.


For realistic particle distributions, the speed up can be much smaller,
especially since a few particles (0.1\% in the 100,000 particle King
model) end up on timesteps smaller than the single stepping timestep.
The speedup of the multistep run over the single step run as calculated
from the above formulae is about a factor of 4.  Note that we have
neglected some costs such as particle pushing and the 
inefficiency of calculating forces for a few particles on modern
pipelined processors.  The actual
factor in wall clock time is 2.7.

\begin{figure}[tb]
\psfig{figure=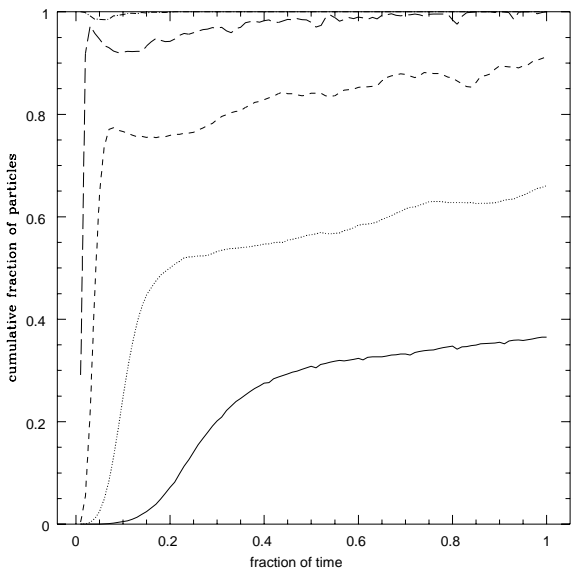,width=\hsize}
\caption{The cumulative fraction of particles on a given timestep or
  larger is
  plotted as a fraction of the age of the Universe in a cosmological
  simulation.  The lowest line is for the longest timestep taken, and
  each upper line is for a factor of two smaller timestep.
\label{cosmostep}
}
\end{figure}

For non-equilibrium situations such as cosmological simulations, a
multistep scheme offers additional speedups since it can adapt to the
changing time scales in the simulation.  An example of the timestep
evolution is shown in Figure \ref{cosmostep}.  Here, the fraction of
particles at a particular timestep or larger is plotted as a function
of time.  Note how almost all particles start out on a very small
timestep at the beginning of the simulation when the Universe is very
dense, and generally migrate to larger timesteps as the simulation
progresses.  One could make adjustments for this trend in a single
step simulation by transforming to an appropriate variable, \eg\ 
expansion factor instead of time (as in Efstathiou \etal\ 1985), but as
can be seen from the figure,
a simple monotonic transformation would not capture all the
complexities of the changing time scales.  Furthermore, one would need
different transformations for different cosmological models.
Finally, to keep the integrations in canonical coordinates would require
that the gravitational softening length is fixed in coordinates
that were different from either comoving or physical (see Appendix A).

From our pragmatic standpoint as users of N-body simulations to model
complex phenomena,
the gain from symplectic integrators is their ability to tolerate
greater  truncation errors in numerical simulations of Hamiltonian systems.  This,
in turn, allows for longer timesteps and shortens the computer time needed to
complete the simulation.  Similarly, individual timesteps in a
particle simulation allows one to concentrate the computational effort
on those particles with the shortest dynamical times that require it.
Either of these is  beneficial only if their advantages 
outweigh their cost, whether in
computational effort or algorithmic complexity.

For a system like the Solar System that is simulated for
billions of dynamical times, a symplectic integrator has obvious advantages.
Any dissipation introduced by truncation error has dire consequences
that are easily observable---a planet may spiral into the Sun.  
The alternatives to symplectic integrators
are either very high order integrators or extremely
short timesteps, or both.  Even a computationally expensive symplectic
integrator can be an overall winner against these
alternatives. (Wisdom and Holman, 1991; Saha and Tremaine, 1992.)

For other systems, such as galaxies or large scale structure, the
situation is not so clear.  These systems are simulated for at most a
few hundred dynamical times, and drifts in conserved quantities due to
truncation error will not be as noticeable.  In these marginal cases, a
symplectic integrator will only be beneficial if it is simple to
implement and computationally cheap.  In this paper we focus on a
criterion based on the local density for this very reason: the
algorithm for estimating the local density is very fast compared to
calculating gravity.  Unfortunately, this criterion is not ideal,
particularly for cosmological simulations.

Is there a better criterion?  If we use the kick-drift-kick form of
leapfrog, one could imagine a timestep criterion that depends only on
the velocity, and therefore commutes with the drift operator.  An
example would be $\tau < \eta \epsilon/v$, where $\epsilon$ is the
force softening.  Three
objections can immediately be raised against such a criterion.  First, it
is not Galilean invariant.  A particle moving with respect to the
coordinate system of the simulation will be on a different timestep
than one that is at rest irrespective of the forces acting on the
particle.  Second, if the acceleration of a particle is nearly opposite to
its velocity at the beginning of a timestep, then the timestep can be
chosen such that the velocity is nearly zero at the middle of the
timestep where the {\tt Select} decision is made.  Such particles will
thereby be always put on a very long timestep.  Third, the choice of
$\epsilon$ to set the scale assumes that the dynamical times are set
by two-body encounters.   For a particle orbiting in a cluster, it is
the cluster density rather than two-body encounters that set the
dynamical time.  On the other hand, in a cluster with significant
substructure the timestep should be set by the encounters with this
substructure.  In this case, $\epsilon/v$ might be an accurate
estimate of the timestep.

An obvious criterion is the acceleration, for example $\tau < \eta
\sqrt{\epsilon/a}$.  This commutes with the {\tt Kick} operator because
{\tt Kick} changes only the velocities, and the acceleration is only a
function of the positions.  So one could easily use it as the select
operator in $DSKD$ just like we use the density criterion.  However, the
acceleration is just the thing that we are trying to minimize
calculating.  Using it as a timestep criterion will cancel the
computational savings we are trying to achieve with multistepping.
For an $SDKD$ method, this is not a problem since one could use the
last calculation of the accelerations that were used to advance the
velocities.  This luxury is not available for the $DSKD$ method, and
one would have to evaluate the acceleration of a particle many times
during the {\tt Select} phase. One could also imagine criteria based on
higher order terms, such as the magnitude of the divergence of the
acceleration, or the
time derivative of acceleration.  However, these tend to be even more
expensive to calculate than the acceleration.

Although our $DSKD$ method 
appears to be symplectic 
when tested with single particle integrations, it is not the
best integrator for general N-body problems.  We simply do not
have a suitable {\tt Select} operator that commutes 
with either the {\tt Drift} or {\tt Kick}
operator, that is computationally cheap to evaluate compared to
an acceleration, and that returns an accurate timestep in all cases.
The local density criterion does well on the first count
but only occasionally satisfies the second.
This problem was foreshadowed by the inability to define the local density
in the Kepler case and having to resort to a global criterion
that used the enclosed
density.  When calculated in a cosmological simulation by averaging
over the nearest $N$ particles, it erased substructure with $\lesssim N$
particles. We have not 
exhausted the search for an ideal timestep criterion, and there yet
may be something that fits these stringent requirements.
Nevertheless, the criterion we have explored in this paper may be very
useful in simulations where structure is defined by large numbers of
particles, such as in the modeling of galactic structure.

For more general cosmological N-body simulations we advocate either
the $SDKD$ or $SKDK$ method using $\eta\sqrt{\epsilon/a}$ and/or (if
clusters have significant substructure) $\eta(\epsilon/v)$ to choose
the timestep.  Using tests similar to those described for our
symplectic integrator we determined that an appropriate choice for
$\eta$ is 0.3 (for Plummer softening 0.4) to insure stability of the
integration.  Although this algorithm is not symplectic, it appears to
give accurate results for quantities of interest in a cosmological
N-body simulation.

This work was supported in part by NASA HPCC/ESS grant NAG 5-2213 and
by NASA Astrophysics Theory grant NAGW-2523.  We also wish to
acknowledge useful discussions with Scott Tremaine.

\appendix
\section{Symplectic Integrators in Comoving Coordinates}

The equations of motion in a comoving coordinate frame are
traditionally presented as
$$
\begin{array}{l}
\dot \vprime + 2 H(t) \vprime = \displaystyle - {\nabla^\prime
\phi^\prime \over a^3} \\
\dot \rprime = \vprime \\
{\nabla^\prime}^2 \phi^\prime = 4 \pi G (\rho^\prime - \rho^\prime_b),
\end{array}
$$
where the primed quantities refer to the comoving coordinate frame,
$a$ is the expansion factor,
$H$ is Hubble's constant, and $\rho^\prime_b$ is the mean background
density.  These can be integrated using the difference equations
$$
\begin{array}{ll}
\rprime_{n+1/2} & = \rprime_n + {1 \over 2} \tau \vprime_n \\
\vprime_{n+1} & = \displaystyle \vprime_n {1 - H(t)\tau \over 1 + H(t)\tau} +
{\nabla^\prime \phi^\prime(\rprime_{n+1/2}) \tau \over a^3 (1+
H(t)\tau)} \\
\rprime_{n+1} & = \rprime_{n+1/2} + {1 \over 2} \tau
\vprime_{n+1}.
\end{array}
$$
(See Hockney and Eastwood, 1981.)  However, these equations do not
appear to be symplectic, \ie, there is no known generating function
that produces this transformation.

However, by making a suitable canonical transformation, one can easily
derive an integrator that is symplectic.   The Lagrangian for the
particle motion in the comoving frame is
$$
{\cal L} = \textstyle{1 \over 2} (a\vprime + \dot a \rprime)^2 - \phi.
$$
With the (time dependent) generating function $\psi = (1/2) a \dot a
\rprime^2$, this transforms to
$$
{\cal L} = \textstyle{1 \over 2}  a^2 \vprime^2 - \phi^\prime/a,
$$
where
$$
\phi^\prime = a\phi + \textstyle{1 \over 2}  \ddot a a^2 \rprime^2.
$$
(See Peebles, 1980.)

Switching to the Hamiltonian formalism,
the momentum canonical to $\rprime$ is $\pprime =
a^2\vprime$, and the Hamiltonian is
$$
H = {\pprime^2 \over 2 a^2 } + {\phi^\prime \over a}.
$$
Although this Hamiltonian is time-dependent\footnote{An energy can
still be defined, as in the Layzer-Irvine energy equation.  See
Peebles, 1980.},
it is separable, so the
``drift'' and ``kick'' operators are easily derived as:
$$
\begin{array}{l}
D(\tau) \equiv \rprime_{t + \tau} = \rprime_{t} + \pprime \displaystyle
\int_t^{t + \tau} {dt \over a^2} \\
K(\tau) \equiv \pprime_{t + \tau} = \pprime_t - \nabla^\prime
\phi^\prime \displaystyle \int_t^{t+\tau} {dt \over a},
\end{array}
$$
where $\tau$ is the step size.
Note that it is assumed here that there is no explicit time dependence in
$\phi^\prime$; \eg\ any softening of the potential must be constant in
comoving coordinates.
For standard cosmologies the integrals in the above operators can be
easily evaluated.  For example in a critical density universe, 
$$
\int_t^{t + \tau} {dt \over a^2} = {2 \over H_0} \left[ a^{-1/2}(t) -
a^{-1/2}(t + \tau) \right],
$$
and
$$
\int_t^{t + \tau} {dt \over a} = {2 \over H_0} \left[ a^{1/2}(t + \tau) -
a^{1/2}(t) \right].
$$
For non-flat matter dominated universes, it is convenient to use the
parametric solutions (Peebles, 1980),
$$
\vcenter{\openup1\jot \halign{$\hfil#$&&${}#\hfil$&\qquad$\hfil#$\cr
a & = A( 1 - \cos \eta ), & t & = B(\eta -\sin \eta ); & \Omega & > 1 \cr
a & = A( \cosh \eta  - 1), & t & = B(\sinh \eta  - \eta); & \Omega & < 1, \cr
}}
$$
where the constants $A$ and $B$ are
$$
A = 4/3 \pi G \rho_b a^3 |R|^2, \quad B = A | R |,
$$
$R$ is the curvature radius and the integrals are
$$
\int_t^{t + \tau} {dt \over a} = {B \over A} \left[ \eta(t + \tau) -
\eta(t) \right],
$$
and
$$
\vcenter{\openup1\jot \halign{$\hfil#$&&${}#\hfil$&\qquad$\hfil#$\cr
\displaystyle{\int_t^{t + \tau} {dt \over a^2}} & \displaystyle{= {B \over A^2} \left[ \cot
{\eta(t) \over 2 } - \cot {\eta(t + \tau) \over 2} \right];}
& \Omega & > 1 \cr
& \displaystyle{= {B \over A^2} \left[ \coth
{\eta(t) \over 2 } - \coth {\eta(t + \tau) \over 2} \right];}
& \Omega & < 1. \cr
}}
$$
The above operators can be used to build leapfrog or higher order
symplectic integrators for particle motion in comoving coordinates.

\end{document}